\documentclass[%
 reprint,
 amsmath,amssymb,
 aps,
 floatfix,
]{revtex4-2}

\usepackage{graphicx}
\usepackage{dcolumn}
\usepackage{bm}

\begin{document}

\preprint{APS/123-QED}

\title{Adversarial attacks on voter model dynamics in complex networks}

\author{Katsumi Chiyomaru}%
\affiliation{%
 Department of Bioscience and Bioinformatics, Kyushu Institute of Technology, Iizuka, Fukuoka 820-8502, Japan
}%
\author{Kazuhiro Takemoto}%
\email{takemoto@bio.kyutech.ac.jp}
\affiliation{%
 Department of Bioscience and Bioinformatics, Kyushu Institute of Technology, Iizuka, Fukuoka 820-8502, Japan
}%

\date{\today}

\begin{abstract}
This study investigates adversarial attacks conducted to distort voter model dynamics in complex networks.
Specifically, a simple adversarial attack method is proposed to hold the state of opinions of an individual closer to the target state in the voter model dynamics. This indicates that even when one opinion is the majority, the vote outcome can be inverted (i.e., the outcome can lean toward the other opinion) by adding extremely small (hard-to-detect) perturbations strategically generated in social networks.
Adversarial attacks are relatively more effective in complex (large and dense) networks.
These results indicate that opinion dynamics can be unknowingly distorted.
\end{abstract}

\maketitle

\section{Introduction}
Opinion dynamics and collective decision-making indicate processes that lead to either a consensus, in which all individuals have the same opinion, or coexistence through competition between different opinions within a population.
These processes were theoretically investigated using the voter model \cite{RevModPhys.81.591,1524860,PhysRevLett.112.158701}.
Voter model dynamics have primarily been investigated for regular lattices.
However, with the development of network science \cite{doi:10.1098/rsta.2012.0375,doi:10.1126/science.1173299} revealing nontrivial connectivity patterns (e.g., small-world topology \cite{i2003-00490-0,PhysRevE.67.035102} and heterogeneous or scale-free connectivity \cite{PhysRevLett.94.178701,PhysRevE.72.036132}) in complex real-world networks, the effects of such patterns on voter model dynamics have also been evaluated.

Given that social networks can influence opinion dynamics because they constrain the flow of information among individuals \cite{doi:10.1126/science.aap9559}, voter model dynamics in complex networks are useful for understanding how to distort collective decision making (e.g., how to disrupt public discourse and democratic decision making) by considering the social network structure.
A previous study \cite{Stewart2019} demonstrated that information gerrymandering, i.e., a specific network connectivity, which indicates who connects to whom, can allow the vote outcome to lean toward one opinion, even when the size of the population with each opinion under the initial state (or each party size) is equivalent and all individuals have the same influence.
In addition, zealots can distort the opinion dynamics \cite{PhysRevE.92.042812,PhysRevE.97.012310}.
A previous study \cite{Stewart2019} showed that a small number of zealots and automated bots can induce information gerrymandering when strategically established in a network.
The results indicate that a vulnerability occurs in which the restricted information flow systematically distorts the collective decision making.

However, alterations to network connectivity, zealots, and bots may be relatively easy to detect; thus, several strategies (e.g., removing zealots and bots and prohibiting alterations in the connectivity) can be considered to avoid such vulnerabilities.
Nevertheless, the adequacy of such defense strategies has yet to be verified.
A different possibility for distorting opinion dynamics can be considered given the analogy between the voter model in complex and neural networks (e.g., the fact that perceptrons in neural networks can be regarded as voters because their outputs are determined by the majority vote \cite{PhysRevE.55.3257}).
Because neural networks are known to be vulnerable to adversarial perturbations (specifically small, i.e., hard-to-detect, perturbations distort their outputs) 
\cite{DBLP:journals/corr/SzegedyZSBEGF13,DBLP:journals/corr/GoodfellowSS14,Yuan2019}, it can be hypothesized that such perturbations can also be generated to distort opinion dynamics in social networks.

In this study, inspired by adversarial attacks on neural network tasks, a simple adversarial attack method is proposed for distorting the voter model dynamics in complex networks and numerically evaluating whether, when one opinion is the majority, the vote outcomes can be shifted toward the other opinion by adding extremely small strategically generated perturbations to social networks.
Evaluations were conducted using models and real-world social networks.
Moreover, the effects of the network size, average node degree, and network connectivity patterns on the outcomes of adversarial attacks were investigated and discussed.

\section{Voter model}
In this study, voter model dynamics in a network with $N$ nodes (individuals) are considered \cite{RevModPhys.81.591,1524860,PhysRevLett.94.178701,PhysRevE.72.036132,PhysRevE.77.041121}.
Each node has one of two discrete opinions at time $t$: $x_i(t)=\{-1,+1\}$ for $i=1,\dots,N$.
Let $\rho_{\mathrm{init}}\in (0,1)$ be the proportion of individuals with opinion $+1$ in the network at time zero. The voter model dynamics start from an initial state in which
the opinions $+1$ are assigned to randomly selected $\rho_{\mathrm{init}}\times N$ nodes, and opinions $-1$ are assigned to the remaining nodes.

For $i=1,\dots,N$, the time evolution of $x_i(t)$ can be described as
\begin{eqnarray}
    x_i(t+1)=
\left\{ 
\begin{array}{ll}
-x_i(t) & \mathrm{with} \ p_i \\
x_i(t) & \mathrm{with} \ 1-p_i
\label{eq:voter_model}
\end{array} \right..
\end{eqnarray}
Note that one (global) time step indicates $N$ node updates.
In Eq. (\ref{eq:voter_model}), $p_i$ is the probability that the opinion of node $i$ is flipped at the next time step (i.e., $t+1$) and is written as
\begin{equation}
    p_i = \frac{1}{2}\left(1-\frac{x_i(t)}{\sum_{h=1}^NA_{ih}}\sum_{j=1}^N A_{ij} x_j(t)\right).
    \label{eq:flip_probability}
\end{equation}
Here, $A_{ij}$ is an element of the weighted adjacency matrix $\boldsymbol{A}$ of the network.
The link weight $A_{ij}$ ($>0$) indicates the influence of neighbor $j$ on individual $i$.
Equations (\ref{eq:voter_model}) and (\ref{eq:flip_probability}) indicate that $x_i(t+1)$ is likely to be the opinion of the majority of neighbors (after considering the link weights) at time $t$. 
For simplicity, complex networks, in which bidirectional links are drawn between individuals and all link weights have a value of 1 (i.e., each individual has the same influence on each neighbor), are considered; that is, $A_{ij}=A_{ji}=1$ if a relationship exists between nodes $i$ and $j$, and $A_{ij}=A_{ji}=0$, otherwise.
Here, self-loops are considered to represent a self-intention: $A_{ii}=1$ for $i=1,\dots,N$.

Equations (\ref{eq:voter_model}) and (\ref{eq:flip_probability}) are computed until the time step reaches $t_{\max}$, and the proportion $\rho$ of individuals with opinion $+1$ is computed as
\begin{equation}
    \rho = \frac{1}{N}\sum_{i=1}^N\delta(x_i(t_{\max}),+1),
    \nonumber
\end{equation}
where $\delta(i,j)$ represents the Kronecker delta.
Notably, $1-\rho$ indicates the proportion of individuals with an opinion $-1$ at time $t_{\max}$.

\section{Adversarial attacks}
Adversarial attacks that distort the voter model dynamics in complex networks consider holding the state of opinions of the individuals at the next time step (i.e., $t+1$) closer to the target state.
Let $x_i^*=\{-1,+1\}$ be the opinion of node $i$ in its target state. The attacks consider making $x_i(t+1)=x_i^*$ for $i=1,\dots,N$ as much as possible.
Because the voter model always reaches a consensus on one opinion in a finite network \cite{RevModPhys.81.591}, adversarial attacks reaching a consensus on opinion $+1$ ($-1$) are considered in this study, specifically, $x_i^{*}=+1~(-1)$ for $i=1,\dots,N$.
The attacks are applied by minimizing the energy $E$ (the negative value of the correlation coefficient between the observed opinion state and target opinion state), which is defined as
\begin{equation}
    E=-\frac{1}{N}\sum_{i=1}^N x_i^*x_i(t+1).
    \nonumber
\end{equation}

We consider minimizing $E$ by temporarily altering the link weights (i.e., by modifying $\boldsymbol{A}$ at each time step); specifically, a perturbation is added to the adjacency matrix at each time step using a gradient descent.
Assuming that the link weights for self-loops and node pairs not connected in the original network are unchangeable, the link weights for node pairs $i$ and $j$, for which $A_{ij}\neq 0$ and $i\neq j$, are perturbed at time $t$ as follows:
\begin{equation}
    A_{ij}^{*}(t) =  A_{ij} - \epsilon \frac{\partial E}{\partial A_{ij}},
    \nonumber
\end{equation}
where $\epsilon$ is a small, positive value.

However, the gradient $\partial E/ \partial A_{ij}$ is not obtained directly (analytically) from the stochastic process described in Eqs. (\ref{eq:voter_model}) and (\ref{eq:flip_probability}).
Thus, we consider a mean-field time evolution of the stochastic process from time $t$ to $t+1$:
\begin{eqnarray}
x_i(t+1)& = & p_i\times -x_i(t) + (1-p_i)\times x_i(t) \nonumber
\\ 
& = & \frac{x_i(t)^2}{\sum_{h=1}^N A_{ih}}\sum_{j=1}^N A_{ij}x_j(t).
\label{eq:MF-time_evolution}
\end{eqnarray}
The gradient $\partial E/ \partial A_{ij}$ in Eq. (\ref{eq:MF-time_evolution}) can be expressed as follows:
\begin{equation}
    \frac{\partial E}{\partial A_{ij}} = -\frac{1}{N}\frac{x_i^*x_i(t)^2}{\left(\sum_{h=1}^N A_{ih}\right)^2}\sum_{\substack{h=1 \\ h\neq j}}^N A_{ih}\left[x_j(t)-x_h(t)\right].
    \label{eq:original_gradient}
\end{equation}


However, the direct use of Eq. (\ref{eq:original_gradient}) may not be useful for adversarial attacks as the perturbation strength is uncontrollable (i.e., high perturbation may be obtained depending on the value of the gradient $\partial E/ \partial A_{ij}$), and computing the sums is costly (e.g., $\sum_{h\neq j} A_{ih}\left[x_j(t)-x_h(t)\right]$).

To avoid these limitations, inspired by the fast gradient sign method \cite{DBLP:journals/corr/GoodfellowSS14} for adversarial attacks on neural network tasks, an optimal maximum-norm constrained perturbation is considered.
Specifically, each element in $\boldsymbol{A}$ is perturbed based on the sign of its gradient 
\begin{equation}
   A_{ij}^{\mathrm{adv}}(t) = A_{ij} - \epsilon \times \mathrm{sign}\left(\frac{\partial E}{\partial A_{ij}}\right), \nonumber
\end{equation}
where $\epsilon$ denotes the strength of perturbation.


From Eq. (\ref{eq:original_gradient}), $\mathrm{sign}(\partial E/ \partial A_{ij}) = -x_i^{*}x_j(t)$ can be estimated because $N>0$, $\sum_{h=1}^N A_{ih}>0$, and $x_i^*$, $x_j(t)$, $x_h(t) = \{+1,-1\}$.
In addition, $x_i(t)^2=1$.
$\mathrm{sign}\left(\sum_{h\neq j} A_{ih}\left[x_j(t)-x_h(t)\right]\right) = {\cal F} = x_j(t)$ because $x_j(t) - x_h(t) = 2x_j(t)$ if $x_j(t) \neq x_h(t)$, and $0$, otherwise.
Note that ${\cal F}=0$ (as a result, $\mathrm{sign}(\partial E/ \partial A_{ij}) = 0$) when all nodes connecting to node $i$ have the same opinion (i.e., $\sum_{h\neq j} A_{ih}\left[x_j(t)-x_h(t)\right]=0$).
However, for simplicity, we can consider that ${\cal F}=x_j(t)$ (as a result, $\mathrm{sign}(\partial E/ \partial A_{ij}) = -x_i^{*}x_j(t)$) in this case because this consideration does not affect the probability $p_i$, when the link weights only for node pairs connected in the original network are perturbed.
In particular, $p_i=0$ whether perturbation is added or not.
Thus, if $A_{ij}\neq 0$ and $i\neq j$, $A_{ij}$ are finally perturbed as follows:
\begin{equation}
    A_{ij}^{\mathrm{adv}}(t)= A_{ij} + \epsilon x_i^{*}x_j(t).
    \label{eq:perturbed_Aij}
\end{equation}
Here, $\boldsymbol{A}^{\mathrm{adv}}(t)$ may be asymmetric.

Adversarial attacks on the voter model dynamics described in Eqs. (\ref{eq:voter_model}) and (\ref{eq:flip_probability}) are conducted using Eq. (\ref{eq:perturbed_Aij}), where $\boldsymbol{A}^{\mathrm{adv}}(t)$ is used at each time step $t$ instead of $\boldsymbol{A}$ in Eq. (\ref{eq:flip_probability}).

Equation (\ref{eq:perturbed_Aij}) indicates the simple adversarial attack strategy, that is, the strengthening of the link weight from node $j$ to node $i$ (i.e., the influence of node $j$ on node $i$) (weakened) if the opinion of neighbor $j$ under the observed state is consistent (inconsistent) with that of node $i$ under the target state (i.e., $x^*_i$).

To compare the performance of the adversarial attacks with random controls, random attacks on the voter model dynamics were considered. Specifically, the adjacency matrix was perturbed as $A_{ij}^{\mathrm{rnd}}(t) = A_{ij} + \epsilon \times s $ for $i$ and $j$, for which $A_{ij}\neq 0$ and $i\neq j$, where $s$ is a random variable uniformly sampled from the set $\{-1,+1\}$.
Random attacks were conducted using $\boldsymbol{A}^{\mathrm{rnd}}(t)$ at each time step $t$ instead of $\boldsymbol{A}$ in Eq. (\ref{eq:flip_probability}).

\section{Simulations}
Supposing that $\rho_{\mathrm{init}}=0.8$ (i.e., the opinion $+1$ is the majority under the initial state), adversarial attacks reaching a consensus on opinion $-1$ (i.e., the target state in which $x_i^{*}=-1$ for $i=1,\dots,N$) are considered.

The adversarial attacks on the voter model dynamics are simulated in complex networks generated from three representative network models: the Erd\H{o}s--R\'enyi (ER) \cite{RevModPhys.74.47,Takemoto2012_book}, Watts--Strogatz (WS) \cite{watts_collective_1998}, and Barab\'asi--Albert (BA) model \cite{RevModPhys.74.47,doi:10.1126/science.286.5439.509}.

ER is a well-used network model that generates random networks by drawing links between $L$ node pairs that are randomly selected from a set of all possible node pairs.
The node degree follows a Poisson distribution with mean $\langle k \rangle=2L/N$.
However, real-world social networks exhibit a non-random structure; they have highly clustered subnetworks and heterogeneous (power-law like) degree distributions \cite{RevModPhys.74.47,Takemoto2012_book}.
Thus, WS and BA models were considered.
By randomly rewiring the links in a one-dimensional lattice, in which each node has $k~(=\langle k \rangle)$ neighbors, with probability $p_{\mathrm{WS}}$, the WS model generates small-world networks whose clustering coefficients are higher than those expected from ER networks.
In this study, $p_{\mathrm{WS}}=0.05$ according to \cite{i2003-00490-0,120006649510}.
In addition, by connecting a newly added node at each time step to $m$ existing nodes using the preferential attachment mechanism, the BA model generates scale-free random networks in which the degree distribution $P(k)$ follows a power law ($P(k)\propto k^{-3}$).
Note that $\langle k \rangle=2m$ for $N \gg 0$.

The voter model dynamics are applied with $t_{\max} = N$ unless otherwise noted because the average time (using the global time step as the unit of measurement) to reach a consensus (consensus time $\tau$) in uncorrelated networks is scaled by at most $N$ \cite{PhysRevE.77.041121}.
The distribution of $\rho$ is obtained from 3000 realizations of the voter model dynamics; moreover, their mean $\langle \rho \rangle$ is computed.

\begin{figure}[htbp]
\includegraphics[width=75mm]{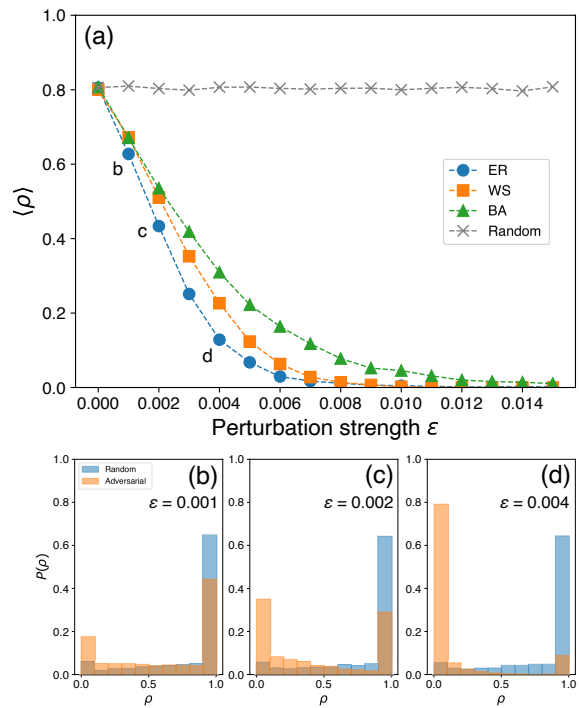}
\caption{\label{fig:eps_vs_rho} (a) Line plot of perturbation strength $\epsilon$ versus $\langle \rho \rangle$. The filled-in symbol indicates $\langle \rho \rangle$ for adversarial attacks against model networks with $N=1000$ and $\langle k \rangle=6$.
A cross indicates $\langle \rho \rangle$ for random attacks against ER networks.
Relative frequency distributions of $\rho$, $P(\rho)$, at $\epsilon$ = (b) 0.001, (c) 0.002, and (d) 0.004 in ER networks.}
\end{figure}

\begin{figure}[htbp]
\includegraphics[width=75mm]{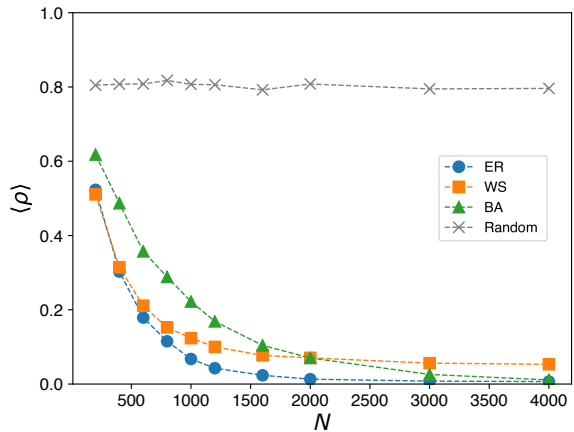}
\caption{\label{fig:N_vs_rho} Line plot of network size $N$ versus $\langle \rho \rangle$. A filled-in symbol indicates $\langle \rho \rangle$ for adversarial attacks with $\epsilon=0.005$ against model networks with $\langle k \rangle=6$. A cross indicates $\langle \rho \rangle$ for random attacks against the ER networks.}
\end{figure}

\begin{figure}[htbp]
\includegraphics[width=75mm]{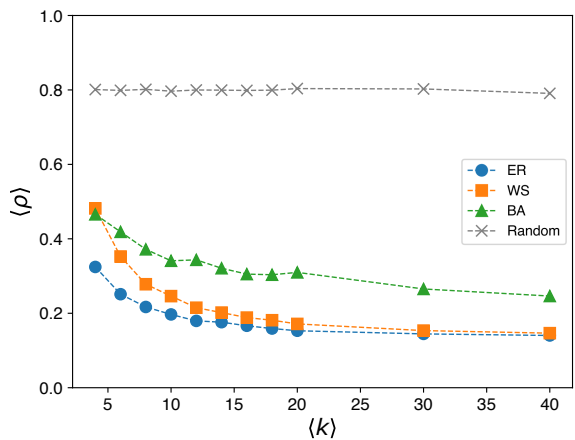}
\caption{\label{fig:kave_vs_rho} Line plot of average degree $\langle k \rangle$ versus $\langle \rho \rangle$. A filled-in symbol indicates $\langle \rho \rangle$ for adversarial attacks with $\epsilon=0.003$ against model networks with $N=1000$.
A cross indicates $\langle \rho \rangle$ for random attacks against ER networks.}
\end{figure}

Figure \ref{fig:eps_vs_rho}(a) shows that $\langle \rho \rangle$ rapidly decreases with the perturbation strength $\epsilon$ for adversarial attacks despite a low $\epsilon$ ($<0.01$).
However, the values $\langle \rho \rangle$ are independent of $\epsilon$ for random attacks (random controls) and are the same as the value at $\epsilon=0$ (i.e., $\langle \rho \rangle$ in the case of no perturbations).
Note that, although the values of $\langle \rho \rangle$ obtained from random attacks are only displayed for the ER networks shown in Fig. \ref{fig:eps_vs_rho}(a), they are also the same for the WS and BA networks (this tendency is similar in Figs. \ref{fig:N_vs_rho} and \ref{fig:kave_vs_rho}).
This indicates that the rapid decrease observed in $\langle \rho \rangle$ with $\epsilon$ results from the adversarial attacks.

Figures \ref{fig:eps_vs_rho}(b)--\ref{fig:eps_vs_rho}(d) show that the distribution of $\rho$ is dramatically changed for adversarial attacks, whereas it is not altered for random attacks. In particular, the figures show the transition from the state under which opinion $+1$ is the majority to the state under which opinion $-1$ is the majority owing to an adversarial attack.
Note that $P(\rho)$ in the case of no perturbations (i.e., at $\epsilon=0$) is similar to that under a random attack (not displayed here to avoid redundancy).

The robustness against adversarial attacks differs slightly among the network models.
For WS and BA networks, in comparison to ER networks, a larger $\epsilon$ is required to decrease $\langle \rho \rangle$ to a desired value owing to an adversarial attack.
However, $\langle \rho \rangle \approx 0$ at $\epsilon = 0.01$ for all model networks, indicating that small perturbations can invert the vote outcomes.

Remarkably, adversarial attacks are more effective for larger networks (Fig. \ref{fig:N_vs_rho}).
For a fixed $\epsilon~(=0.005)$, $\langle \rho \rangle$ rapidly decreases with network size $N$ for an adversarial attack, whereas it is independent of $N$ for random attacks.
Note that all voter model dynamics are applied with $t_{\max} = 1000$, thereby demonstrating that the observed $N$-dependency on $\langle \rho \rangle$ is independent of $t_{\max}$. However, a similar tendency (i.e., a rapid decrease in $\langle \rho \rangle$ with $N$) is also observed when $t_{\max} = N$.

Adversarial attacks are also more effective for denser networks (Fig. \ref{fig:kave_vs_rho}).
For a fixed $\epsilon~(=0.003)$, $\langle \rho \rangle$ decreases with the average degree $\langle k \rangle$ for an adversarial attack but is independent of $\langle \rho \rangle$ for a random attack.
However, the effect of $\langle k \rangle$ on the decrease in $\langle \rho \rangle$ is less remarkable than the effect of $N$ (Fig. \ref{fig:N_vs_rho}).

\begin{figure}[htbp]
\includegraphics[width=82mm]{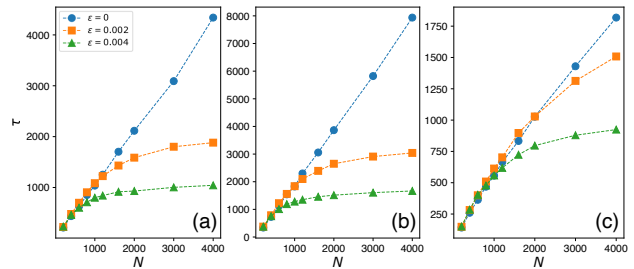}
\caption{\label{fig:N_vs_consensus_time} Line plot of network size $N$ versus consensus time $\tau$ in (a) ER, (b) WS, and (c) BA networks with $\langle k \rangle=6$.
Here, $\tau$ is obtained from 3000 realizations of the voter model dynamics.
}
\end{figure}

Adversarial attacks shorten the consensus time $\tau$ despite a small $\epsilon$ (Fig. \ref{fig:N_vs_consensus_time}).
This tendency is remarkable for a large $N$; specifically, $\tau$ increases in a sublinear manner with $N$ compared to the case without perturbations (i.e., $\epsilon=0$).
However, for a relatively small $N$, adversarial attacks may require a slightly longer $\tau$ compared to the case of $\epsilon=0$.
This is because the consensus state is antagonistic between opinions $+1$ and $-1$ owing to the weak effect of adversarial attacks for a relatively small $N$ (Fig. \ref{fig:N_vs_rho}).

\begin{figure}[htbp]
\includegraphics[width=75mm]{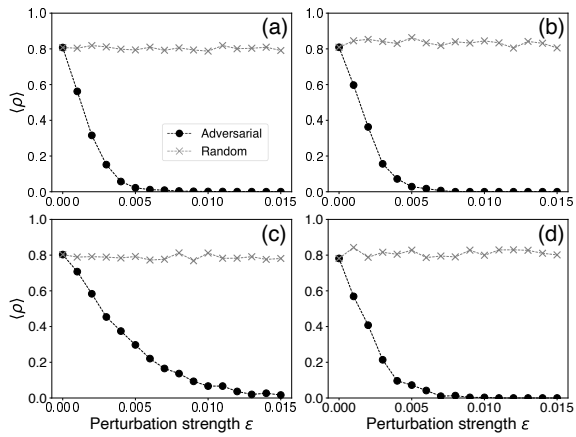}
\caption{\label{fig:eps_vs_rho_real} Line plots of perturbation strength $\epsilon$ versus $\langle \rho \rangle$ in real-world social networks:
(a) Facebook ($N=4039$ and $\langle k \rangle = 43.7$),
(b) Advogato ($N=5054$ and $\langle k \rangle = 16.6$),
(c) AnyBeat ($N=12645$ and $\langle k \rangle = 7.8$),
and (d) HAMSTERster ($N=2000$ and $\langle k \rangle = 16.1$).}
\end{figure}

Adversarial attacks on real-world social networks were also investigated (Fig. \ref{fig:eps_vs_rho_real}).
Facebook \cite{NIPS2012_7a614fd0}, Advogato \cite{NetworkDataRepository,massa2009bowling}, AnyBeat \cite{NetworkDataRepository,Fire2012}, and HAMSTERster \cite{NetworkDataRepository} networks were considered.
These networks are undirected.
For simplicity, the largest connected component in each real-world network was used, and all link weights were set to 1.
Voter model dynamics were applied with $t_{\max}=1000$ for each network; moreover, $\langle \rho \rangle$ was obtained from 300 realizations.
As shown in Fig. \ref{fig:eps_vs_rho_real}, $\langle \rho \rangle$ rapidly decreases with $\epsilon$ for adversarial attacks, whereas it is independent of $\epsilon$ and is the same value under no perturbations for random attacks. 
These results indicate that a small perturbation ($\epsilon < 0.01$) can invert the vote outcomes.
A simple comparison of the adversarial robustness (i.e., the minimum $\epsilon$ required to decrease $\langle \rho \rangle$ to the desired $\langle \rho \rangle$) between networks is inaccurate because $N$ and $\langle k \rangle$ differ.

\section{Discussion}
Herein, it is demonstrated that the voter model dynamics in both the model and real-world complex networks can be distorted by adding extremely small perturbations to the networks (link weights in particular) using the proposed adversarial attack method (Figs. \ref{fig:eps_vs_rho} and \ref{fig:eps_vs_rho_real}).
Previous studies have considered the introduction of relatively easy-to-detect perturbations such as zealots and alterations to network connectivity to distort the opinion dynamics in complex networks; however, this study shows that imperceptible (hard-to-detect) perturbations can distort such dynamics.
Because link weights are interpreted as the contact frequencies between individuals, perturbations against link weights indicate a change in contact frequencies.
For example, in social networking services, it may be easy to change such contact frequencies by manipulating the display frequencies of the other posts of individuals on the timeline of each individual. Adversaries who have data on social networks and the opinions of individuals and can modify the contact frequencies may then be able to control the opinion dynamics (and the subsequent vote outcomes) by slightly increasing or decreasing the display frequencies based on the opinions of the individuals.
Adversarial attacks (Eq. \ref{eq:perturbed_Aij}) are easy to implement.
Importantly, social networks remain mostly unchanged even when adversarial attacks are conducted.
Moreover, zealots and automated bots are not required.
It is possible for opinion dynamics to become distorted.

Larger and denser networks are more vulnerable to adversarial attacks (Figs. \ref{fig:N_vs_rho} and \ref{fig:kave_vs_rho}).
Further investigation is needed for a deeper mathematical explanation (e.g., using mean-field approximation approaches \cite{PhysRevE.77.041121,Carro2016,Huang_2017}, stochastic pair approximation\cite{Peralta_2018}, and approximate master equations \cite{PhysRevX.3.021004}). This vulnerability occurs owing to the flip probability (Eq. (\ref{eq:flip_probability})) being changed through perturbations.
For simplicity, supposing a network in which every node has neighbors with the same number of opinions $+1$ and $-1$, adversarial attacks toward the node having the target opinion $-1$ are considered.
Given Eq. (\ref{eq:flip_probability}), for each node, the flip probability is $1/2$ with no perturbations; however, it increases (decreases) by $\epsilon/2$ with perturbations if the opinion is $+1$ ($-1$).
Although this change appears to be minor for each node, it significantly affects the dynamics of the entire network.
For example, the probability that all nodes will have the target opinion at the next time step (i.e., the probability that all nodes with opinion $-1$ at time $t$ will also have opinion $-1$ at time $t+1$ and all nodes that have opinion $+1$ at time $t$ will have opinion $-1$ at time $t+1$) is $(1/2)^{N}$ with no perturbations; however, the value increases to $[1-(1-\epsilon)/2]^{N^{(-)}}[(1+\epsilon)/2]^{N^{(+)}}=[(1+\epsilon)/2]^N$ when perturbations occur, where $N^{(-)}$ and $N^{(+)}$ are the numbers of nodes with opinions $-1$ and $+1$ at time $t$, respectively.
In brief, the probability increases $(1+\epsilon)^N$ times when perturbations occur as compared to the presence of no perturbations.
Therefore, adversarial attacks can distort the voter model dynamics with a small perturbation; moreover, they are more advantageous for larger networks.
Similarly, adversarial attacks can reduce the consensus time (Fig. \ref{fig:N_vs_consensus_time}).
In addition, adversarial attacks are more effective when perturbations are added to all possible node pairs because the influence of individuals with the target opinion can be utilized, although for greater realism, adding perturbations is limited to only connected node pairs in this study.
Therefore, adversarial attacks are advantageous for dense networks.

Given the results shown in Figs. \ref{fig:eps_vs_rho}--\ref{fig:kave_vs_rho}, a heterogeneous connectivity and small-world topology may weakly inhibit adversarial attacks.
In heterogeneous networks, when one opinion is the majority and hubs have the same opinion, the existence of the hubs inhibits adversarial attacks because the hubs affect the opinions of other individuals; in addition, their opinions are relatively stable even if the opinions of a few individuals are changed through an adversarial attack.
A small-world topology inhibits the ordering process of voter model dynamics \cite{i2003-00490-0,PhysRevE.67.035102}; thus, it may also inhibit adversarial attacks.
It would be interesting to determine a type of network structure that will enhance or inhibit an adversarial attack.

Given that the voter model dynamics are approximated with the mean-field time evolution to estimate the gradient $\partial E/ \partial A_{ij}$, the proposed method may be ineffective for specific networks, although it was confirmed to be useful as a representative model (Fig. \ref{fig:eps_vs_rho}) and for several real-world (Fig. \ref{fig:eps_vs_rho_real}) networks.
In this context, to conduct more effective adversarial attacks in future investigations, it would be interesting to improve the proposed method and propose novel methods using different approaches.
Furthermore, methods for applying adversarial attacks should be more sparsely developed.
Although the proposed method is simple and effective, it requires an adjustment of the link weights in the network.

The adversarial attacks considered in this study are limited to complex networks in which the relationships between individuals are bidirectional, and all link weights are the same.
Thus, it would also be interesting to further investigate adversarial attacks against complex networks in which the relationships between individuals are asymmetric \cite{PhysRevE.81.057103}, where the link weights vary (i.e., each individual has a different influence on each neighbor) \cite{PhysRevE.83.066117}. The connections are temporally altered (e.g., forming relationships among individuals of similar beliefs \cite{PhysRevE.74.056108}), and there are several types of relationships \cite{Amato2017}.
Moreover, adversarial attacks should be evaluated using more realistic voter models (e.g., noisy voter \cite{Carro2016} and game-theoretic voter \cite{Stewart2019} models) and real-world experiments (as in \cite{Stewart2019}).

Thus, adversarial attacks on opinion dynamics in complex networks will become a new line of research.

The data and relevant code for this research are stored in the author’s GitHub repository \cite{github_advvoter}. 

\begin{acknowledgments}
This study was supported by JSPS KAKENHI (Grant Number 21H03545).
We would like to thank Editage (www.editage.jp) for providing the English language editing.
\end{acknowledgments}

\bibliography{apssamp}

\end{document}